\documentclass{PoS}
\usepackage{amsmath} 
\usepackage{amsfonts} 
\usepackage{amssymb}
\newcommand{\real}{\mathrm{Re}}
\newcommand{\imag}{\mathrm{Im}}
\renewcommand{\d}{\mathrm{d}}
\newcommand{\T}{\mathrm{T}}

\newcommand{\Z}{\mathbb{Z}}

\newcommand{\J}{\mathbb{J}}

\newcommand{\trace}{\mathrm{Tr}}

\title{Simulating low dimensional QCD with Lefschetz thimbles}

\ShortTitle{Simulating lattice QCD with Lefschetz thibles}

\author{\speaker{Christian Schmidt}\\
        Universit\"at Bielefeld, Fakult\"at f\"ur Physik, D-33615 Bielefeld, Germany\\
        E-mail: \email{schmidt@physik.uni-bielefeld.de}}

\author{Felix Ziesch\'e\\
        Universit\"at Bielefeld, Fakult\"at f\"ur Physik, D-33615 Bielefeld, Germany\\
        E-mail: \email{fziesche@physik.uni-bielefeld.de}}

\abstract{Non-perturbative lattice QCD calculations at non vanishing
  baryon number density are hampered by the QCD sign problem. The path
  integral, that in lattice QCD is calculated numerically, becomes
  highly oscillating. One possible solution is the Lefschetz thimble
  approach. It requires a deformation of the original integration
  domain into a manifold embedded in complex space. For properly
  chosen integration manifolds (``thimbles'') the sign problem is
  drastically alleviated. For some bosonic and fermionic models this
  approach has been shown to work. Here we apply the thimble
  disretization to $(0+1)$-dimensional QCD with standard staggerd
  quarks and disscuss issues that may arrise in higher dimensions.}

\FullConference{34th annual International Symposium on Lattice Field Theory\\
		24-30 July 2016\\
		University of Southampton, UK}

\begin{document}

\section{Introduction}
Lattice QCD, which has proven to be extremely successful at zero
baryon number density, is hampered by the infamous sign problem once a
nonzero baryon chemical potential ($\mu_B$) is introduced. The reason
lies in the complex phase that is acquired by the fermion determinant
at $\mu_B>0$. The fermion determinant, which is obtained after
integrating out the Grassmann valued fermion fields from the QCD
partition function is part of the weight in lattice QCD
calculations. It is subject to importance sampling during a Monte
Carlo integration of the grand canonical partition function. On a
Euclidean hyper-cubic lattice one finds
\begin{equation}
Z(T,V,\mu_B)=\int \prod_{x,\nu}{\rm d} U_{x,\nu}\;
{\rm det[M(U_{x,\nu},\mu_B)]}e^{-S_G(U_{x,\nu})}
\equiv\int \prod_{x,\nu}{\rm d} U_{x,\nu}\;
e^{-S_{\rm eff}(U_{x,\nu},\mu_B)}\;,
\label{eq:Z}
\end{equation}
where $S_G$ denotes the Euclidean gauge action, $x$ labels the
space-time points and $\nu$ the space-time directions. $U_{x,\nu}$ are
the $\mathrm{SU}(3)$-valued link variables, that are associated with
the gauge fields in the continuum. The integral has dimension
$n=8\times 4 \times V$, where $V$ is the number of lattice points. As
soon as the determinant is not a real and positive function anymore,
standard MC methods break down.  The so-called ``sign problem'' arises
from the fluctuating sign of the determinant that leads to extreme
cancellations and an exponentially suppressed signal-to-noise ratio.

In the past, many attempts have been made to circumvent this problem,
including reweighting \cite{Fodor:2001au}, Taylor expansion in the
chemical potential \cite{Allton:2002zi, Gavai:2003mf}, analytic
continuation from purely imaginary chemical potentials
\cite{deForcrand:2002hgr, D'Elia:2002gd}, canonical partition
functions \cite{Alexandru:2005ix, Kratochvila:2005mk} and strong
coupling methods \cite{Kloiber:2013rba, deForcrand:2014tha}.  However,
they all have certain limitations. Two of the most recent strategies
are based on a complexification of the integration variables
\cite{Sexty:2013ica, Cristoforetti:2012su}. In case of the link
variables, one exploits Gell-Mann's representation
$U_{x,\mu}=\exp\{-\frac{i}{2}\sum_a\omega_{x,\nu}^{a}\lambda^{a}\}$,
where $\lambda_a$ denote the Gell-Mann matrices. The complexification
is performed by allowing the real parameters $\omega_{x,\nu}^a$ to
become generally complex, {\it i.e.}  the link variables become
elements of $\mathrm{SL}(3,\mathbb{C})$.  A new idea recently put
forward is to deform the original domain of integration from the real
hyperplane to some submanifold of the complex space.  Picard-Lefschetz
theory tells us that associated with each complex saddle point
$\sigma$ of the effective action $S_{\rm eff}$ as introduced in
Eq.~(\ref{eq:Z}), there exists a manifold $\J_\sigma$ (``thimble'')
such that the integral $Z$ can be written as a linear combination of
Integrals over the thimbles,
\begin{equation}
Z=\sum_\sigma n_\sigma I_\sigma,\quad \mbox{with}\quad 
I_\sigma=\int\limits_{\J_\sigma} \prod_{x,\nu}{\rm d} 
U_{x,\nu}\;e^{-S_{\rm eff}(U_{x,\nu},\mu_B)}
\quad \mbox{and}\quad n_\sigma\in\Z.
\label{eq:thimbles}
\end{equation}
The thimble manifolds exhibit the appealing feature that the imaginary
part of $S_{\rm eff}$ is constant for all points on the thimble, the
integrals $I_\sigma$ are thus non oscillating and can thus be
performed numerically. Motivated by Witten's work \cite{Witten:2010cx,
  Witten:2010zr}, this approach has been recently applied to various
bosonic and fermionic models, for a recent review see
\cite{Scorzato:2015qts}.

\section{QCD in $(0+1)$-dimensions with diagonal Polyakov loops}
It is well known that the QCD partition function in $(0+1)$-dim.
simplifies drastically. Due to the low dimensionality there is no
gauge action and we are left with the fermion determinant only, which
can be expressed as a determinant over a reduced matrix $M^{\rm red}$
that depends on the Polyakov loop ($P$) and anti-Polyakov loop
($P^{-1}$). The partition function can thus be expressed as
\cite{Bilic}
\begin{equation}
Z^{(N_f)}=\int \mathrm{d}P \; \mathrm{det}^{N_f}[M^\mathrm{red}]\;, 
\quad \mbox{with}\quad M^\mathrm{red}=2\cosh(\mu_c/T)\mathbf{1}
+e^{\mu/T}P+e^{-\mu/T}P^{-1}\;,
\label{eq:Z1d}
\end{equation}
where $\mathrm{d}P$ is the Haar measure over the gauge group
$\mathrm{SU}(3)$. Here $\mu_c$ is the critical chemical potential,
given in terms of the quark mass ($m$) as
\begin{equation}
a\mu_c=\mathrm{arcsinh}(am)\;,
\end{equation}
with $a$ being the lattice constant. In the following we will restrict
ourselfs to the one flavor partition function ($N_f=1$).

\begin{figure}[t]
\begin{center}
\begin{minipage}[t]{0.30\textwidth}
\includegraphics[width=\textwidth]{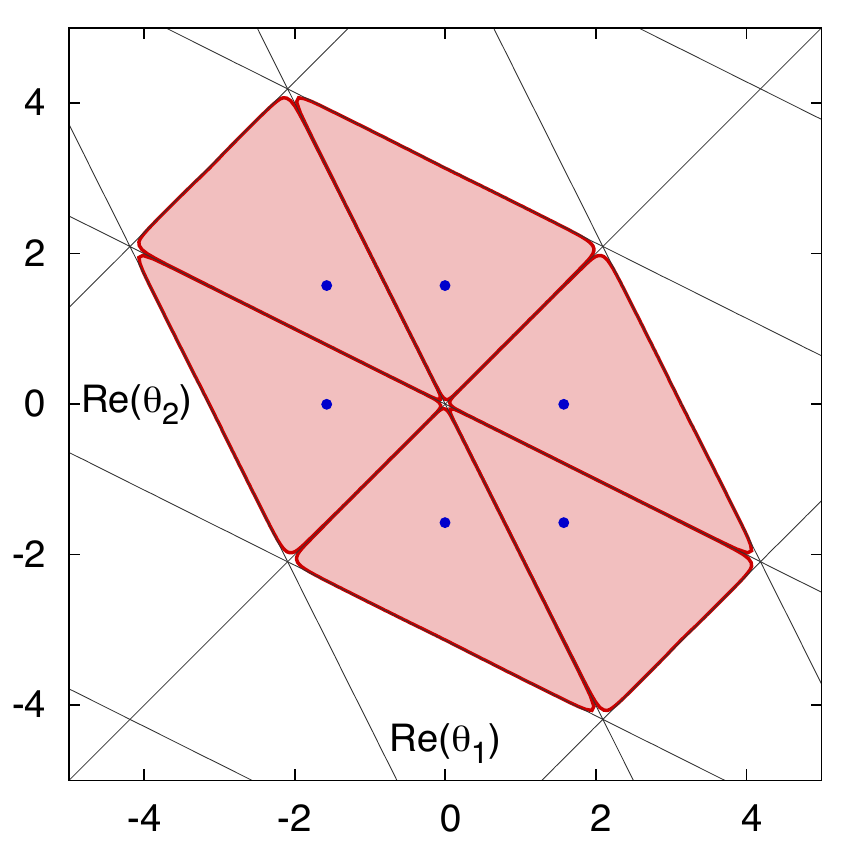}
\end{minipage}
\begin{minipage}[t]{0.315\textwidth}
\includegraphics[width=\textwidth]{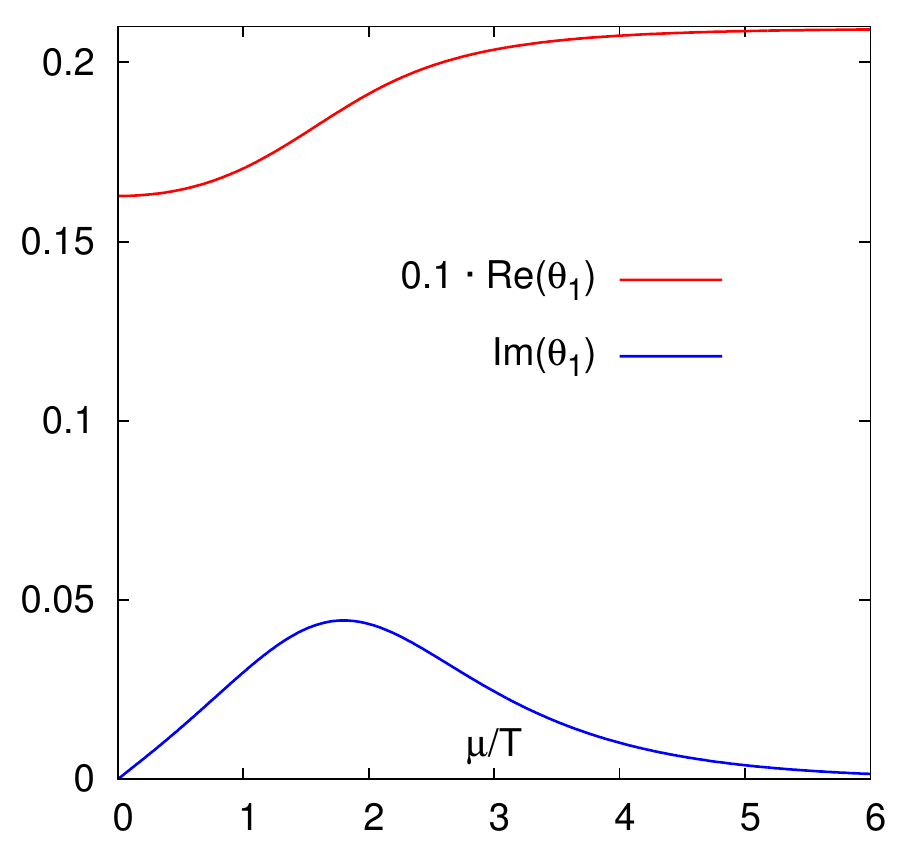}
\end{minipage}
\begin{minipage}[t]{0.32\textwidth}
\includegraphics[width=\textwidth]{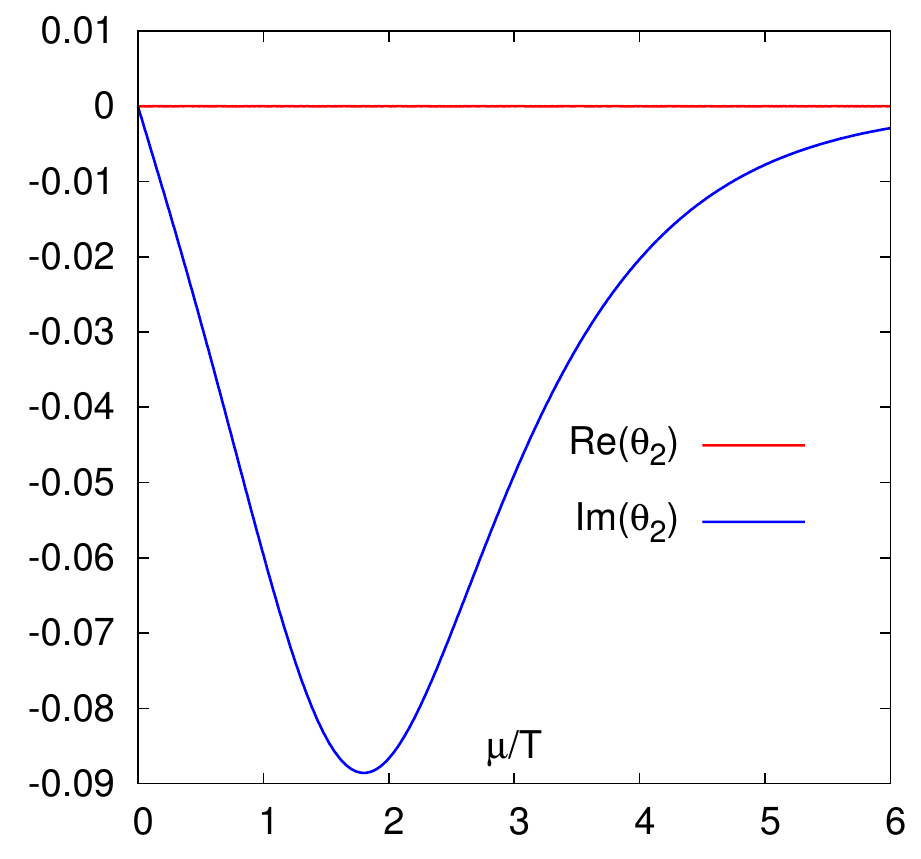}
\end{minipage}
\caption{The Weyl chambers of the reduced Haar measure (left). For the
  first six chambers the saddle points at $\mu=0$ are indicated as dots. The
  dependence of the coordinates of the saddle points on the chemical
  potential are shown in the middle and right panel. }
\label{fig:saddles}
\end{center}
\end{figure}

Before we start dealing with $\mathrm{SL}(3,\mathbb{C})$ valued
integration variables, we observe that in this $(0+1)$-dim. case the
Polyakov loop can be diagonalized. On the price of introducing an
additional term, the Jacobian ($J_H$), we can parametrize the Polyakov
loop in terms of 2 phases $\theta_1$ and $\theta_2$ as
\begin{equation}
P=\mathrm{diag}(e^{i\theta_1},e^{i\theta_2},e^{-i(\theta_1+\theta_2)})\;.
\end{equation}
The Jaccobian is therefore given as
\begin{equation}
J_H(\theta_1,\theta_2)=
\frac{8}{3\pi^2}
\sin^2\left(\frac{\theta_1-\theta_2}{2}\right)
\sin^2\left(\frac{2\theta_1+\theta_2}{2}\right)
\sin^2\left(\frac{\theta_1+2\theta_2}{2}\right)\;.
\end{equation}
As long as we are interested in the partition function itself, or
observables that are invariant under transformations $P\to
P'=UPU^{-1}$, it suffices to generate diagonal Polyakov loops and the
partition function (\ref{eq:Z1d}) can be written as integral over the
phases $\theta_1$ and $\theta_2$ as
\begin{equation}
Z=\int \mathrm{d}\theta_1\mathrm{d}\theta_2 \; 
e^{-S_\mathrm{eff}(\theta_1,\theta_2)}\;,
\end{equation}
where we define the effective action as 
\begin{eqnarray}
S_\mathrm{eff}&=&-\ln J(\theta_1,\theta_2) 
-\ln\left\{2\cosh(\hat\mu_c)+2\cosh(\hat\mu+i\theta_1)\right\} \nonumber \\
&-&\ln\left\{2\cosh(\hat\mu_c)+2\cosh(\hat\mu+i\theta_2)\right\} 
-\ln\left\{2\cosh(\hat\mu_c)+2\cosh(\hat\mu-i[\theta_1+\theta_2])\right\}\;.
\label{eq:Seff1d}
\end{eqnarray}
Here we denote $\hat\mu=\mu/T$. At zero $\mu$, this effective
action diverges on the grey lines shown in Fig.~\ref{fig:saddles}
which
naturally divide the phase space into separated regions. Inside of
each region we find a single minimum of the effective action, its
location is shown in Fig.~\ref{fig:saddles} (left) as blue dot. The
minima become saddle points of $\real[S_\mathrm{eff}]$ in the
complexified space $(\real[\theta_1],\real[\theta_2],\imag[\theta_1],
\imag[\theta_2])$ of real-dimension 4. The tangent
space of each stable thimble at the critical points is spanned by the
Takagi vectors of the Hessian matrix $\frac{\partial}{\partial
  \theta_i}\frac{\partial}{\partial \theta_j}S_\mathrm{eff}$.
Not surprisingly, at $\mu=0$ each tangent space as well as each thimble
itself is real, and can be identified with each other.  The thimbles
just coincide with the triangular regions shown in
Fig.~\ref{fig:saddles} (left). Due to the symmetries of the reduced
Haar measure it suffices to integrate only over one of the thimbles,
{\it i.e.} each thimble gives the same contribution to the partition
function and they correspond to the Weyl chambers of the reduced Haar 
measure.

At $\mu\ne0$, we observe that the saddle points drift into complex
space but come back to the real plane at very large $\mu$, as
indicated in the middle and right panel of Fig.~\ref{fig:saddles}.
Here the thimbles as well as their tangent spaces at the saddle points
($\T_\sigma\J$) become complex. The thimbles are now also curved
manifolds which can not be identified with their tangent spaces.

\begin{figure}[t]
\begin{center}
\begin{minipage}[t]{0.295\textwidth}
\includegraphics[width=\textwidth]{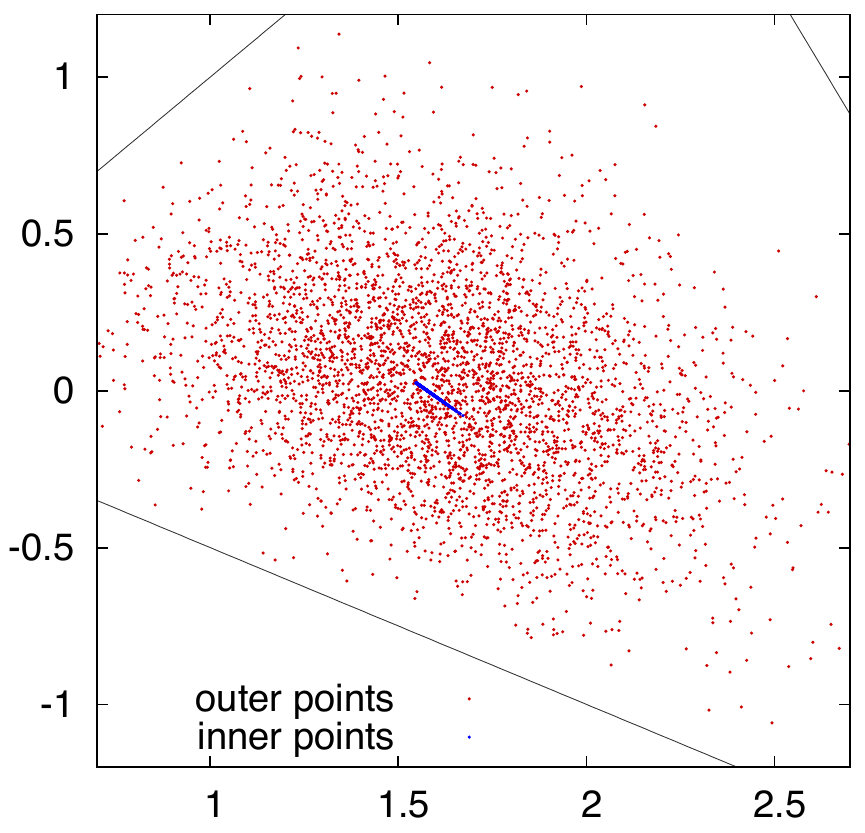}
\end{minipage}
\begin{minipage}[t]{0.315\textwidth}
\includegraphics[width=\textwidth]{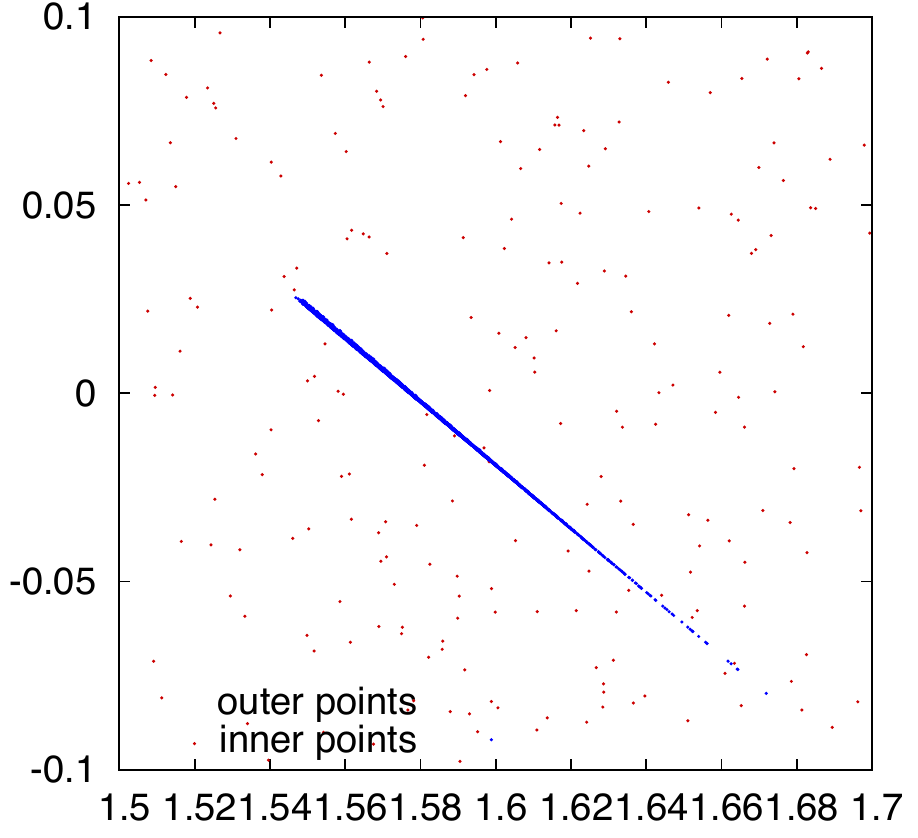}
\end{minipage}
\end{center}
\caption{{\it Right:} Graphical represantation of the contraction map
  $F_\tau^{-1}(\theta)$ for $\mu=0$ and $\tau=0.7$.  Sample points on the
  tangent space of the thimble are depicted in blue, while the
  red points indicate their corresponding images on the thimble. 
  {\it Left:} The same but zoomed in towards the critical piont.}
\label{fig:contraction}
\end{figure}

To sample the thimble integrals (\ref{eq:thimbles}) numerically,
several algorithm have been proposed \cite{Cristoforetti:2012su,
  Fujii:2013sra, Mukherjee:2014hsa, DiRenzo:2015foa,Alexandru:2015xva}. 
We will apply here the contraction algorithm
proposed in \cite{Alexandru:2015xva}. It relies on the fact, that the
thimble $\J$ can be generated from an $\varepsilon$-environment
$U_\varepsilon \subset \T_\sigma\J$ around the saddle point $\sigma$ 
by making use of the flow map $F_\tau(\theta)$ of the initial value 
problem defined by the steepest ascent (SA) equation
\begin{equation}
\frac{\d}{\d t}\theta=\left(\frac{\d S_\mathrm{eff}}{\d\theta}\right)^*, 
\qquad \theta(t_0)\in
U_\varepsilon\;.
\end{equation}
Here $\theta$ denotes a vector of complex-dimension 2. For a fixed
flow time $\tau>t_0$, $F_\tau(\theta):\mathbb{C}^2\to \mathbb{C}^2$
defines a map and in particular we have
$F_\tau(U_\varepsilon)\simeq\J$ if $\varepsilon$ is sufficiently
small. Note that we can not take the limit $\varepsilon \to 0$, as the
image of $F_\tau$ will then be reduced to a single static point, the
critical point $\sigma$. Note also that integrating the SA equation is
numerically stable, thus for large flow times $\tau$, the flow will be
pushed towards the stable thimble $\J$. The idea of the contraction
algorithm is to integrate the thimble by sampling $\T_\sigma\J$, 
{\it e.g.} with a random-walk Metropolis algorithm.  The algorithm is
depicted graphically in Fig.~\ref{fig:contraction}.
Mathematically, this procedure corresponds to a variable
transformation of the integral. Thus, we have to take into account an
additional Jacobian $J_F$, which we obtain by flowing a complete set
of orthonormal basis vectors of the tangent space from the sampling
point to the thimble point and taking the determinant to the linear
map they define. It is clearly visible from Fig.~\ref{fig:contraction}
(right) that an anisotropic proposal probability is advisable, as the
Takagi values of the Hessian are rather different.

The contraction algorithm can be used to calculate one-thimble
expectation values of thermodynamic observables $O$, defined as
\begin{equation}
\left<O\right>_\sigma=\frac{1}{Z_\sigma}\int_{\J_\sigma}\mathrm{d}\theta_1\mathrm{d}\theta_2\; 
O\; e^{-S_\mathrm{eff}(\theta_1,\theta_2)},
\quad \mbox{with}\quad Z_\sigma=\int_{\J_\sigma}\mathrm{d}\theta_1\mathrm{d}\theta_2\; 
e^{-S_\mathrm{eff}(\theta_1,\theta_2)}\;.
\end{equation}
We stress again, that the integral over a single thimble is completely
free of any sign problem stemming from the fermion determinant. There
is, however, a much reduced residual sign problem, which due to the 
fact that the integration manifold is curved and is here taken into account 
through the complex valued Jacobian $J_F$.

It was pointed out in
Ref.~\cite{Alexandru:2015sua} that for short flow times $\tau$, the
image $F_\tau(U_\varepsilon)$ does indeed come close to several
thimbles. This way, the contraction algorithm for small $\tau$, can be
used to obtain expectation values according to the correct partition
function as defined in Eq.~(\ref{eq:Z}).  This version of the
contraction algorithm will be call Maryland algorithm in the
following. Although the sign problem now comes back into the game, as
the phase of the determinant is not constant on the considered manifold
$F_\tau(U_\varepsilon)$, we find that after reweighting with the
phase, the Maryland algorithm yields indeed correct expectation values
for $(0+1)$-dimensional QCD.

\section{QCD in $(0+1)$-dimensions with general Polyakov loops}
\begin{figure}[t]
\label{fig:action}
\begin{center}
\begin{minipage}[t]{0.3\textwidth}
\includegraphics[width=\textwidth]{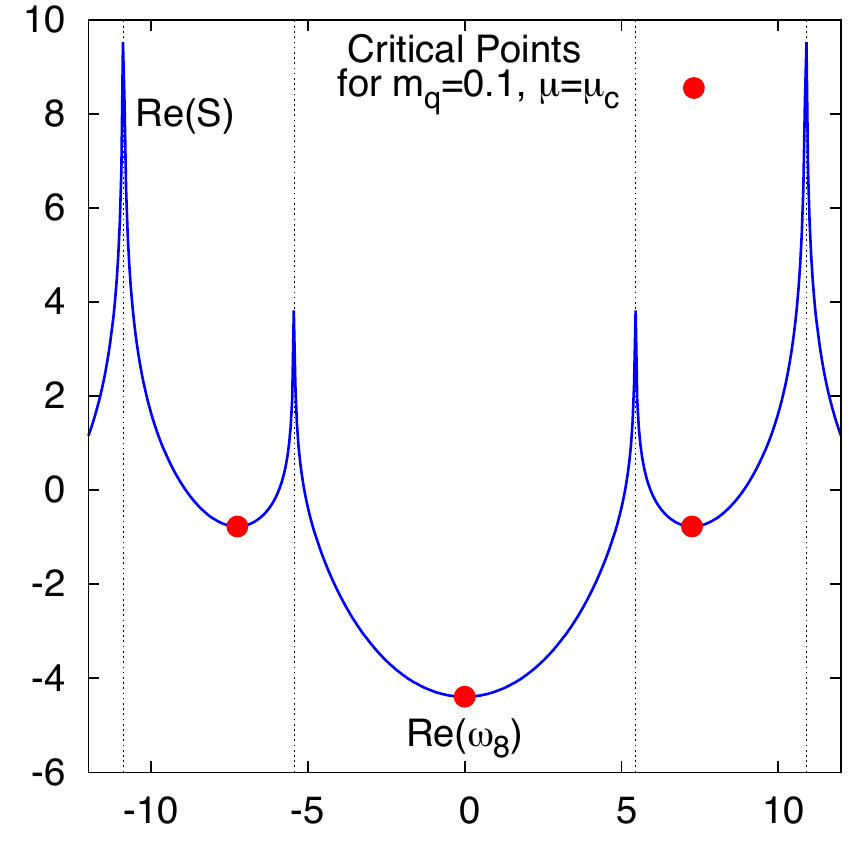}
\end{minipage}
\begin{minipage}[t]{0.293\textwidth}
\includegraphics[width=\textwidth]{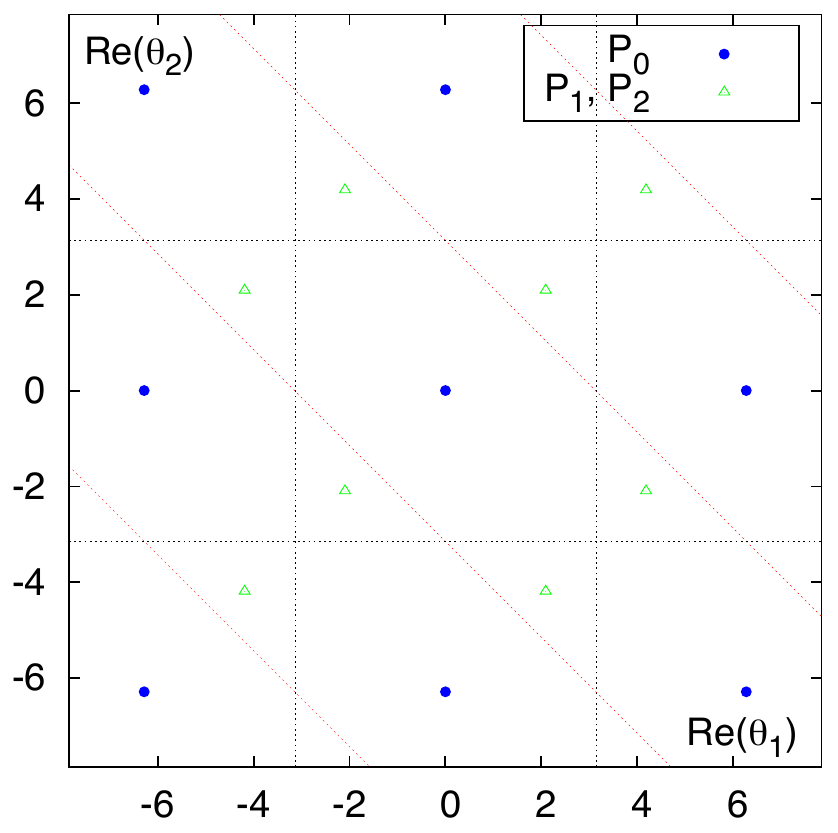}
\end{minipage}
\begin{minipage}[t]{0.31\textwidth}
\includegraphics[width=\textwidth]{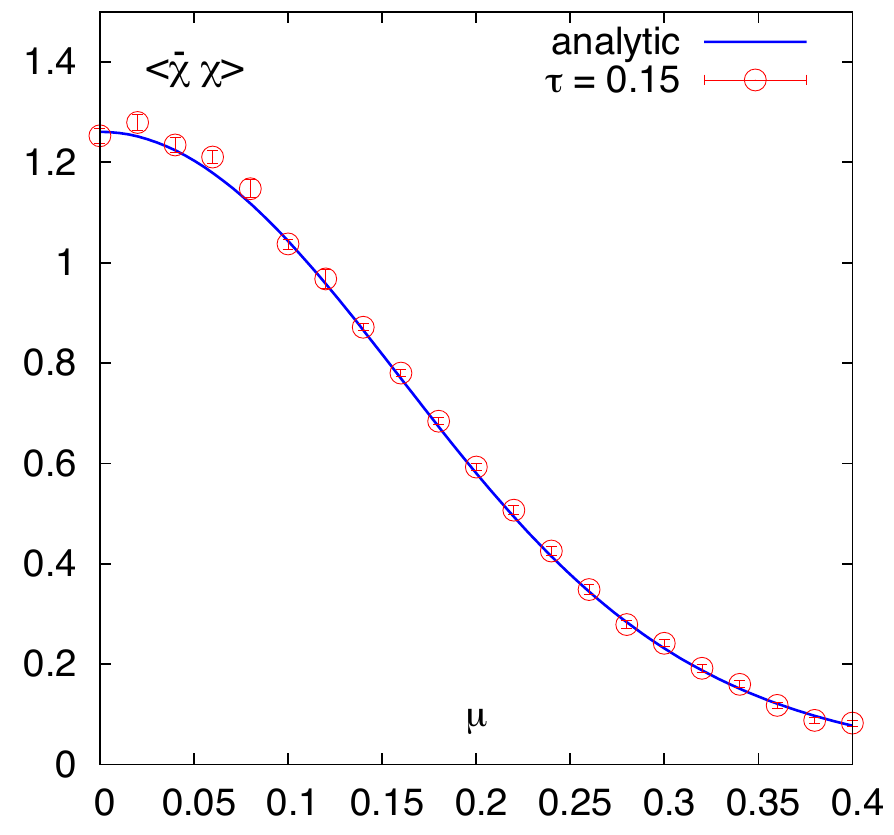}
\end{minipage}
\end{center}
\caption{{\it Left:} Action as function of the parameter $\real[\omega_8]$
  for $\mu=\mu_c$ and $am=0.1$. {\it Middle:} Projection of critical points and thimbles onto the 
  ($\theta_1,\theta_2$)-plane, as defined in the text. {\it Right:} Chiral condensate in
  $(0+1)$-dimensional QCD for $am=0.1$ and $T=1/4a$ as a function of
  $a\mu$. The results have been obtained with the contraction
  algorithm at flow time $\tau=0.15$. Also shown is the exact analytic
  result.}
\end{figure}

We now come back to $\mathrm{SU}(3)$-valued variables and parameterize
the Polyakov loop as
$P=\exp(-\frac{i}{2}\sum_a\omega_a\lambda_a)$. The number of
independent variables is thus increased from 2 to 8, leading to 16
degrees of freedom in the complexified theory. Without the Jacobian of
the reduced Haar measure we find already at $\mu=0$ (up to
periodicity) three inequivalent thimbles at $P=\mathbf{1},
e^{i2\pi/3}\mathbf{1}$ and $ e^{-i2\pi/3}\mathbf{1}$, in
correspondence with the holonomy sectors of the Polyakov loop under
generalized gauge transformations. This structure remains valid for
$\mu\ne0$. The relative weight between the thimbles varies with $\mu$,
whereas the critical points itself appear to be $\mu$-independent.  In
Fig.~\ref{fig:action}(left) we show the action of as function of the
parameter $\real[\omega_8]$ for $\mu=\mu_c$, which features the three
critical points as local minima.
When the thimbles are projected to the same space as shown in
Fig.~\ref{fig:saddles} (left), the plane spanned by the angle
variables $\real[\theta_1]$ and $\real[\theta_2]$, which are now given
as $\theta_1=-\omega_3/2-\omega_8/(2\sqrt{3})$ and $\theta_2=\omega_3/2-\omega_8/(2\sqrt{3})$, we
find that they are bounded by the (projected) lines shown in
Fig.~\ref{fig:action} (middle).
 
As usual the tangent space of the thimbles at the critical points are spanned
by the Takagi vectors of the Hessian, now given as
\begin{equation}
H_{kl}=\frac{\partial^2 S}{\partial\omega_k\partial\omega_l}=\trace\left[ M^{-1}\frac{\partial M}{\partial \omega_l}M^{-1}\frac{\partial M}{\partial \omega_k}
 - M^{-1}\frac{\partial^2 M}{\partial \omega_k \partial \omega_l}\right]\;.
\label{eq:hessian}
\end{equation}
We find that at the critical points the Hessian takes the form
$H=\gamma\mathbf{1}$, with a real valued $\gamma$ for the main thimble
$\J_0$, ($P=\mathbf{1}$), and a general complex $\gamma$ for the two 
others. As a consequence the tangent space at the three critical
points do not coincide but are rotated with respect to each other. We
believe that this fact renders the Maryland algorithm rather
ineffective. Nevertheless, we find that it gives the correct results
as shown in Fig.~\ref{fig:action} (right) on the example of the chiral
condensate.

We note that a more effective algorithm for this simple case of
$(0+1)$-dimensional might be obtained by following the procedure as
proposed in Ref.~\cite{Mukherjee:2013aga}.  If the intersection
$n_\sigma$ numbers are known, which is true in our case where all are
equal to one, we can pick a thimble $\mathbb{J}_\sigma$ with
probability $\frac{n_\sigma}{\sum n_\sigma}$, perform an update there
and count how often the we accept to jump to a different
thimble. This shall give the correct weight $c_k$ for the expectation
value $\left<O\right>_\sigma$, according to
\begin{equation}
\left< O\right> 
= \frac{ \sum_\sigma n_\sigma e^{-i S_I(P_\sigma)} \int_{\mathbb{J}_\sigma} \mathrm{d}P \; O(P)\; e^{-S_R(P)}} 
{\sum_\sigma n_\sigma e^{-iS_I(P_\sigma)}Z_\sigma} 
= \sum_\sigma c_\sigma \left< O \right>_\sigma \;.
\end{equation}

\section{Outlook: QCD in $(1+1)$-dimensions}
Now we have set the stage for $(1+1)$-dimensional QCD, which is work
in progress. It shall be very interesting to see if the three basic
thimbles remain, or if the thimble structure gets considerably
complicated. Unfortunately, a reduced definition of the fermion
matrix, given only in terms of Polyakov loops is no longer
available. It will be desirable to work in temporal gauge (or any
other fixed gauge), which reduces the number of integration variables,
and avoids complications due to gauge orbits. Nevertheless, the
computation of the Hessian, Eq.~(\ref{eq:hessian}), will become
numerically much more involved. Also, for the calculation of the
action, the residual phase \cite{Cristoforetti:2014gsa}, or similarly
the Jacobian in case of the contraction algorithm
\cite{Alexandru:2016lsn}, one eventually needs to find a stochastic
approximation.

\section*{Aknowledgements}
We whish to thank W. Unger and A. Lindemeier for helpfull discussions
and comments.

\end{document}